\documentclass[aps,prb,twocolumn,longbibliography,notitlepage,nofootinbib,superscriptaddress]{revtex4-2}

\usepackage[utf8]{inputenc}
\usepackage{amsmath,amssymb,amsfonts,stmaryrd}
\usepackage{physics}
\usepackage{dsfont} 
\usepackage{graphicx}
\usepackage{xcolor}
\usepackage{graphicx}
\usepackage{cancel}
\usepackage{natbib}
\usepackage{color}
\usepackage{tikz}
\usepackage{mathrsfs}
\usepackage{relsize}
\usepackage{multirow}
\usepackage[linktocpage]{hyperref}
\usepackage[all]{xy}
\hypersetup{colorlinks=true,citecolor=blue,linkcolor=blue, urlcolor=blue, breaklinks=true}

\usepackage[mathscr]{euscript}
\usepackage[scr=boondox]{mathalpha}
\usepackage{comment}

\usepackage{float}

\usepackage{bm} %allows for bold math type
\usepackage{subfigure} %allows for 2x2 figures
\usepackage{environ}
\usepackage{url}
\usepackage[margin=0.75in]{geometry}

\usepackage{mathtools}

\newcommand{\U}{\text{U}(1)}
\newcommand{\Z}{\mathbb{Z}}
\newcommand{\OO}{\text{o}}

\newcommand{\SO}{\mathscr{S}_{\OO}}

\newcommand{\PO}{\vec{\mathscr{P}}_{\OO}}

\begin{document}

\title{Fractionally Quantized Electric Polarization and Discrete Shift of \\ Crystalline Fractional Chern Insulators}

\author{Yuxuan Zhang and Maissam Barkeshli}
\affiliation{
Department of Physics and Joint Quantum Institute,
College Park, Maryland 20742, USA}

\begin{abstract}
Fractional Chern insulators (FCI) with crystalline symmetry possess topological invariants that fundamentally have no analog in continuum fractional quantum Hall (FQH) states. Here we demonstrate through numerical calculations on model wave functions that FCIs possess a fractionally quantized electric polarization, $\vec{\mathscr{P}}_{\OO}$, where $\OO$ is a high symmetry point.
$\vec{\mathscr{P}}_{\OO}$ takes fractional values as compared to the allowed values for integer Chern insulators because of the possibility that anyons carry fractional quantum numbers under lattice translation symmetries. $\vec{\mathscr{P}}_{\OO}$, together with the discrete shift $\mathscr{S}_{\OO}$, determine \it fractionally quantized \rm universal contributions to electric charge in regions containing lattice disclinations, dislocations, boundaries, and/or corners, and which are fractions of the minimal anyon charge. We demonstrate how these invariants can be extracted using Monte Carlo computations on model wave functions with lattice defects for 1/2-Laughlin and 1/3-Laughlin FCIs on the square and honeycomb lattice, respectively, obtained using the parton construction. These results comprise a class of fractionally quantized response properties of topologically ordered states that go beyond the known ones discovered over thirty years ago.
\end{abstract}

\maketitle

The recent experimental discoveries of fractional Chern insulators (FCIs) in two-dimensional materials \cite{Spanton2018FCI,cai2023signatures,lu2024fractional,zeng2023thermodynamic,park2023observation,xu2023observation,xie2024even,lu2024extended} raise a fundamental question of whether these states are adiabatically connected to known continuum fractional quantum Hall (FQH) states. The key distinction for FCIs is the presence of strong lattice effects \cite{kalmeyer1987equivalence,wen1989csl,kol1993fractional,hafezi2007,tang2011high,regnault2011,sun2011,parameswaran2013,neupert2011fractional} and, in the clean limit, crystalline symmetries such as lattice translational and rotational symmetries. The existence of crystalline symmetries implies topological invariants and quantized responses with no continuum analog \cite{barkeshli2012a,manjunath2021cgt,manjunath2020FQH}, implying that FCIs can be in a fundamentally different universality class than continuum FQH states.  

Recently it has been understood how integer Chern insulators with crystalline symmetry possess several many-body topological invariants \cite{zhang2022fractional,zhang2022pol,zhang2023complete,manjunath2023classif}, two of which include a quantized electric polarization $\vec{\mathscr{P}}_\OO$ and the discrete shift $\mathscr{S}_\OO$, where $\OO$ is a high symmetry point in the unit cell \cite{zhang2022fractional,zhang2022pol,zhang2024pol}. These invariants determine universal contributions to the electric charge from lattice dislocations, disclinations, boundaries, and corners. In this paper we demonstrate through systematic numerical calculations on model wave functions for FCIs the existence of a \it fractionally quantized \rm electric polarization and discrete shift. These invariants govern universal contributions to the electric charge from lattice defects that are fractions of the minimal anyon electric charge. 

We note that our predictions may be testable experimentally. It has been demonstrated that dislocation pairs in graphene can be created and moved through controlled focused electron beam irradiation \cite{Warner2012,Lehtinen2013,Quang2015,bonilla2015measuring}. Dislocations also appear in twisted bilayer graphene \cite{deJong2022}, as a dislocation in one of the layers of the twisted bilayer graphene induces a dislocation in the moire lattice \cite{Cosma2014}. While isolated disclinations are not generally energetically stable in two-dimensional electron systems, they can potentially be designed in synthetic quantum systems, such as ultracold atoms, topological photonic systems, and superconducting qubits, where topologically ordered states have been realized experimentally \cite{semeghini2021probing,satzinger2021realizing,iqbal2024non,ozawa2019}.

Let us review the key properties of $\vec{\mathscr{P}}_\OO$ and $\mathscr{S}_\OO$ for integer Chern insulators of fermions with Hall conductivity $\sigma_H = C e^2/h$, where $C$ is the integer Chern number \cite{zhang2022pol,manjunath2021cgt,zhang2024pol}. With lattice (magnetic) translational symmetry and $\mathbb{Z}_M$ lattice rotational symmetry about $\OO$, there is an electric polarization $\vec{\mathscr{P}}_{\OO}$ which is well-defined modulo an integer vector, denoted as $\vec{\mathscr{P}}_{\OO} \mod \mathbb{Z}^2$. The rotational symmetry quantizes the allowed values of $\vec{\mathscr{P}}_{\OO}$, such that $(1 - U(2\pi/M)^T)\vec{\mathscr{P}}_{\OO} \in \mathbb{Z}^2$ is an integer vector, where $U(2\pi/M)$ is a $2\pi/M$ rotation in the lattice basis. This implies $\vec{\mathscr{P}}_{\OO}$ is a $\mathbb{Z}_2 \times \mathbb{Z}_2$, $\mathbb{Z}_3$, $\mathbb{Z}_4$, or $\mathbb{Z}_1$ invariant when $M = 2,3,4,6$. 
On a square lattice, $M = 4$ and $\vec{\mathscr{P}}_{\OO} = (0,0)$ or $(1/2,1/2) \mod \mathbb{Z}^2$. On a honeycomb lattice with $\mathbb{Z}_3$ rotational symmetry, $\vec{\mathscr{P}}_{\OO} = (0,0)$, $(1/3, 2/3)$, or $(2/3, 1/3) \mod \mathbb{Z}^2$. $\mathscr{S}_{\OO}$ is a $\mathbb{Z}_{M}$ invariant:
it is integer or half-integer modulo $M$, with the constraint $\mathscr{S}_{\OO} = \frac{C}{2} \mod 1$.  $\vec{\mathscr{P}}_\OO$ and $\mathscr{S}_\OO$ only depend on the maximal Wyckoff position (MWP) of $\OO$, and this dependence is determined by the charge and magnetic flux per unit cell \cite{zhang2022pol} (see Table \ref{table:shiftingO} in Appendix \ref{app:reviewTorsion}). 

For FCIs, $\vec{\mathscr{P}}_\OO$ and $\mathscr{S}_\OO$ can take fractional values as compared with their counterparts in integer CIs \cite{manjunath2021cgt,manjunath2020FQH}. The quantization rules depend on the fusion rules and fractional braiding statistics of the anyons and will be reviewed in Appendix \ref{app:reviewTorsion}, \ref{app:response}. 
For the $1/2$-Laughlin state of bosons with\footnote{Here and in the rest of the paper we use natural units, $e = \hbar = 1$.} $\sigma_H = \frac{1}{2} \frac{1}{2\pi}$ on a square lattice where the origin $\OO$ is at a 4-fold rotational symmetry point, the allowed values of the polarization can be written as $\vec{\mathscr{P}}_\OO =  \frac{p_\OO}{4} (1,1) \mod \frac{1}{2}\mathbb{Z}^2$, where $p_\OO$ is an integer \cite{manjunath2021cgt}. The case where $p_\OO$ is odd arises because of fractional quantum numbers of the semion under translations, as specified by the ``discrete torsion vector" discovered in \cite{manjunath2021cgt} and reviewed in Appendix~\ref{app:reviewTorsion}, \ref{app:response}. The discrete shift  $\mathscr{S}_\OO = s_\OO/2 \mod 2$, where $s_\OO$ is an integer. The possibility that $\mathscr{S}_{\OO}$ is a half-integer in bosonic systems is fundamentally tied to fractional quantum numbers of the semion under lattice rotations as specified by $s_\OO$. For the $1/3$-Laughlin state of fermions on a honeycomb lattice with $\sigma_H = \frac{1}{3} \frac{1}{2\pi}$ and $\OO$ chosen to be a 3-fold rotational symmetry point ($M_{\OO} = 3$), $\vec{\mathscr{P}}_\OO = \frac{p_\OO}{9} (1,2) \mod \frac{1}{3}\mathbb{Z}^2$ and $\mathscr{S}_{\OO} = s_{\OO}/3 \mod 1$, where now $s_{\OO}$ is a half-integer ($s_{\OO} = 1/2 \mod 1$) and $p_{\OO}$ is a integer. Thus we have three possible values, $\mathscr{S}_{\OO} = 1/6, 1/2, 5/6 \mod 1$ and $\vec{\mathscr{P}}_{\OO} = (0,0), (1/9, 2/9), (2/9, 1/9) \mod \frac{1}{3} \mathbb{Z}^2$.

\section*{Universal charge response of defects}

\begin{figure*}[t]
    \centering
    \includegraphics[width=17.5cm]{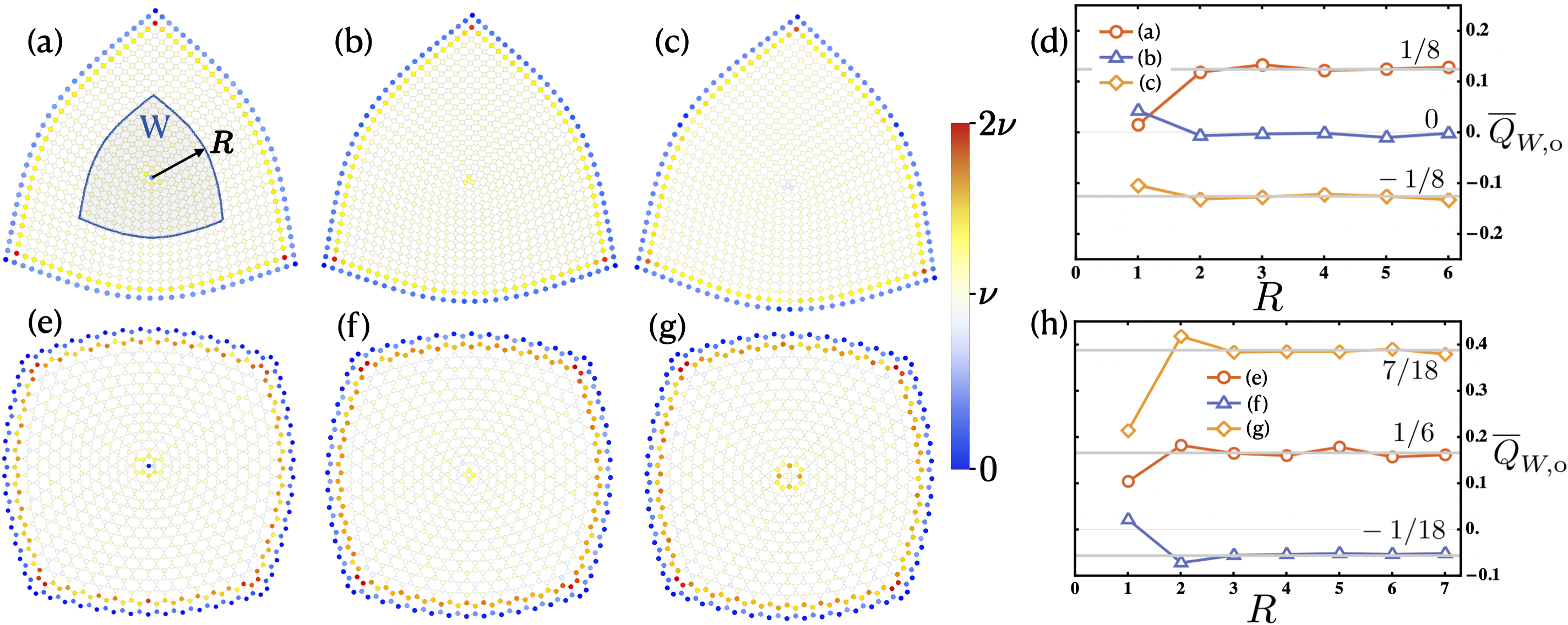}
    \caption{\textbf{(a$\sim$d)}1/2-Laughlin state on a $\Gamma=\pi/2$ disclination, filling 100 fermions in the whole disk.  \textbf{(a)} Charge profile for a vertex-centered disclination in units of $\nu$ for $\OO=\beta$, parton shifts $\mathscr{S}_{\beta,1}=\mathscr{S}_{\beta,2}=1/2$; \textbf{(b)} Plaquette-centered disclination with $\OO=\alpha$, $\mathscr{S}_{\alpha,1}=1/2, \mathscr{S}_{\alpha,2}=7/2$; \textbf{(c)} Plaquette-centered disclination with $\OO=\alpha$, $\mathscr{S}_{\alpha,1}=1/2, \mathscr{S}_{\alpha,2}=5/2$. \textbf{(d)} $\overline{Q}_{W,\OO}$ as a function of radius of $W$. 
    \textbf{(e$\sim$h)}1/3-Laughlin state on a $\Gamma=2\pi/3$ disclination in the honeycomb lattice, filling 85 fermions in the whole disk.  \textbf{(e)} Charge profile of a vertex-centered disclination in units of $\nu$ for $\OO=\beta_1$, $\mathscr{S}_{\beta,1}=\mathscr{S}_{\beta,2}=\mathscr{S}_{\beta,3}=1/2$; \textbf{(f)} Plaquette-centered disclination with $\OO=\alpha$, $\mathscr{S}_{\alpha,1}=1/2, \mathscr{S}_{\alpha,2}=\mathscr{S}_{\alpha,3}=5/2$; \textbf{(g)} Plaquette-centered disclination with $\OO=\alpha$, $\mathscr{S}_{\alpha,1}=1/2, \mathscr{S}_{\alpha,2}=\mathscr{S}_{\alpha,2}=3/2$. \textbf{(h)} $\overline{Q}_{W,\OO}$ as a function of radius of $W$.}
    \label{fig:disclinationCom}
\end{figure*}

The invariants $\vec{\mathscr{P}}_\OO$ and $\mathscr{S}_\OO$ determine universal contributions to the electric charge in subregions of the system. For a given system, pick a choice of unit cell and a spatial region $W$ that is large compared to the correlation length $\xi$ (for example as shown in Fig. \ref{fig:disclinationCom}(a)). The boundary $\partial W$ should lie along the boundary of the unit cells and far away (compared to $\xi$) from the boundary of the system and any lattice dislocations or disclinations. The total charge in $W$ is defined as:
\begin{align}\label{eq:charge_definition}
    Q_W:=\sum_{i\in W}\text{wt}(i)Q_i,
\end{align}
where $\sum_i$ is over all lattice sites $i$ in $W$. $\text{wt}(i)=1$ for the interior points and $\text{wt}(i)=\theta_i/(2\pi)$ if the site $i$ is on the boundary of $W$, where $\theta_i\in[0,2\pi)$ is the angle subtended by the interior of $W$ \cite{zhang2022fractional}. $Q_i$ is the ground state expectation value of the charge on site $i$. 
Let $q_\star$ be the minimal quasiparticle charge. We find that $Q_W \mod q_{\star}$ is universal, independent of perturbations, and can be decomposed into distinct contributions:
\begin{align}
\label{QwEq}
    Q_W = \frac{\Gamma}{2\pi}\SO  + \vec{L}_{\OO}\cdot\vec{\mathscr{P}}_{\OO} + \nu n_{W,\OO}+\sigma_{H}\delta \Phi_{W,\OO}\mod q_{\star}
\end{align}
Here $\Gamma$, $\vec{L}_{\OO}$, and $n_{W,\OO}$ are certain geometrical quantities that depend on the subregion enclosed by $W$, which keep track of corner and disclination angles, vectorized edge lengths, and number of unit cells, respectively. See \cite{zhang2024pol} for a detailed discussion, which we briefly review below.

Consider a loop $\gamma_{\OO}$ in $W$ obtained by a sequence of elementary lattice translations starting at $\OO$. $\gamma_\OO$ is far from any lattice defects and lattice boundaries, and is smoothly deformable to the boundary $\partial W$ without crossing any lattice defects or boundaries. The vectorized length $\vec{L}_{\OO}\equiv \sum_{j\in \gamma_{\OO}}\hat{L}_{j}$ is the sum of unit translation vectors $\hat{L}_{j}\in\{\pm \hat{x},\pm \hat{y}\}$ over $\gamma_{\OO}$. $j$ indicate points on $\gamma_{\OO}$ related by the translations, all $j$ points correspond to the same MWP as $\OO$. 
When $\gamma_{\OO}$ only encloses a dislocation, $\vec{L}_{\OO}$ is the Burgers vector $\vec{b}_{\OO}=\vec{L}_{\OO}$. The term $\vec{L}_{\OO} \cdot \vec{\mathscr{P}}_{\OO}$ thus indicates the universal contribution from lattice dislocations, impure lattice disclinations, and lattice boundaries.

$\Gamma$ is defined as $\Gamma\equiv 2\pi - \sum_{j\in\gamma_{\OO}} K_j$, where $K_j$ is the geodesic curvature of the loop $\gamma_{\OO}$ at $j$. Specifically, $K_j = \pi-\theta_j$ where $\theta_j$ is the angle subtended by the inside of the loop at $j$. We define $\theta_j$ to be in the range $0<\theta_j<2\pi$, which enables us to define $\Gamma$ absolutely instead of just modulo $2\pi$. When $\gamma_{\OO}$ only encloses a disclination, the disclination angle $\Omega_{\text{disc}}=\Gamma$; when $\gamma_{\OO}$ only encloses a boundary with corner(s), the total corner angle $\Omega_{\text{cor}}$ can be determined with $\Omega_{\text{cor}}=\Gamma-2\pi$. Therefore the term $\frac{\Gamma}{2\pi} \mathscr{S}_\OO$ above indicates the universal contribution from lattice disclinations and corners.

\begin{figure}[t]
    \centering
    \includegraphics[width=8.5cm]{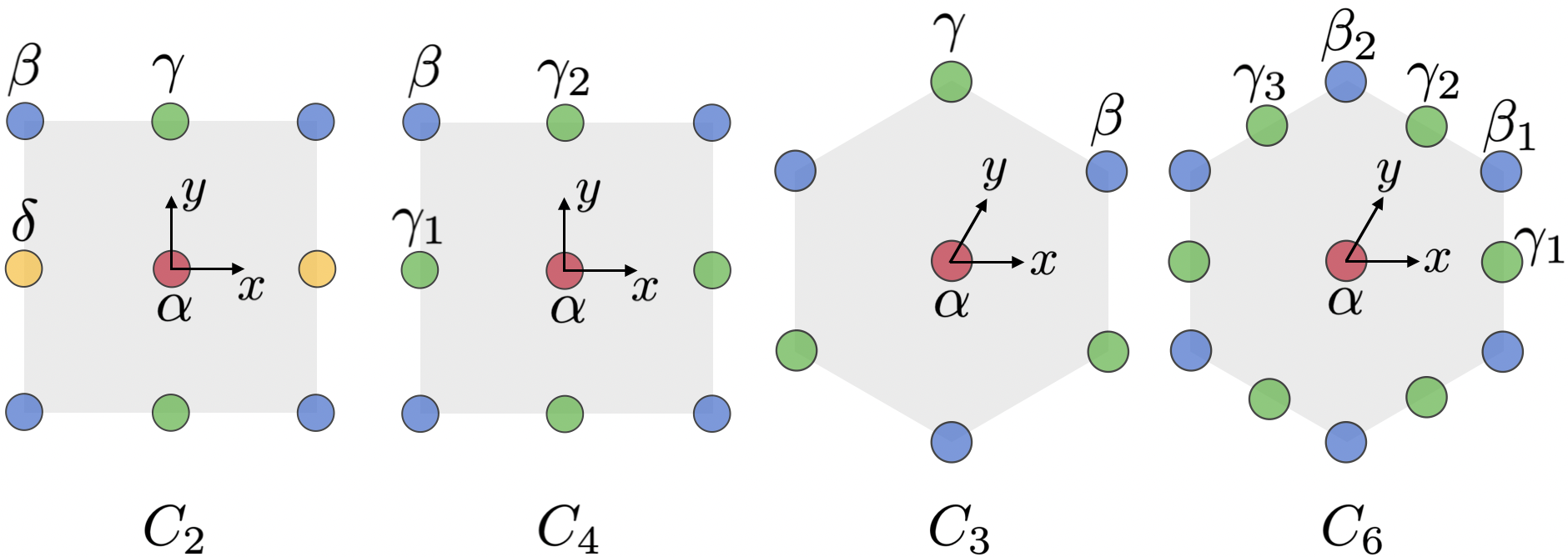}
    \caption{Unit cells for $M=2,3,4,6$. The points $\OO = \alpha, \beta, \gamma, \delta$ are the maximal Wyckoff positions.}
    \label{fig:unit_cells}
\end{figure}

\bgroup
\def\arraystretch{1.4}
\begin{table}
    \centering
    \vspace{0.3cm}
    \begin{tabular}{c|c|c|c|c}
    $M$&$\OO$&$M_{\OO}$&$\vec{n}_{\OO}$& $m_{\OO}$ \\
    \hline
    2&  $\alpha$&2&(0,0)&1\\
    &$\beta$&2&(1/2,1/2)&0\\
    &$\gamma$&2&(1/2,0)&0\\
    &$\delta$&2&(0,1/2)&0\\
    \hline
    4&$\alpha$&4&(0,0)&1\\
    &$\beta$&4&(1/2,1/2)&0\\
    &$\gamma_2$&2&(1/2,0)&0\\
    &$\gamma_1$&2&(0,1/2)&0\\
    \multicolumn{4}{c}{}
    \end{tabular}\qquad
    \begin{tabular}{c|c|c|c|c}
    $M$&$\OO$&$M_{\OO}$&$\vec{n}_{\OO}$& $m_{\OO}$ \\
    \hline

    3&$\alpha$&3&(0,0)&1\\
    &$\beta$&3&(1/3,2/3)&0\\
    &$\gamma$&3&(2/3,1/3)&0\\
    \hline
    6&$\alpha$&6&(0,0)&1\\
    &$\beta_1$&3&(1/3,2/3)&0\\
    &$\beta_2$&3&(2/3,1/3)&0\\
    &$\gamma_1$&2&(0,1/2)&0\\
    &$\gamma_2$&2&(1/2,0)&0\\
    &$\gamma_3$&2&(1/2,1/2)&0\\
    \end{tabular}
\caption{$\vec{n}_{\OO}$ and $m_{\OO}$ for all high symmetry points. $M_\OO$ denotes the order of the point group at $\OO$.}\label{tab:nuc}
\end{table}
\egroup

When $\Gamma \neq 0 \mod 2\pi$, $\vec{L}_{\OO}$ depends on the origin $\OO$.
Under a shift $\OO \rightarrow \OO + \vec{v}$, $\vec{L}_{\OO+\vec{v}}=\vec{L}_{\OO} - (1-U(\Gamma))\vec{v}.$ $U(\Gamma)$ represents a counterclockwise rotation by $\Gamma$. If $\vec{v} = \vec{\Lambda} \in \mathbb{Z}^2$, an integer vector, the MWP of $\OO$ is unchanged under this shift. This implies an equivalence $\vec{L}_{\OO}\simeq\vec{L}_{\OO}-(1-U(\Gamma))\vec{\Lambda}$. Importantly, $(1 - U(\Gamma))\vec{\Lambda} \cdot \vec{\mathscr{P}}_{\OO} = 0 \mod q_\star$ (see Appendix~\ref{app:reviewTorsion}), so that the contribution to $Q_W \mod q_\star$ only depends on the equivalence class of $\vec{L}_{\OO}$. 
On the square lattice for example, for $\OO=\alpha$ or $\beta$ there are two equivalence classes: $\vec{L}_{\OO} \simeq (0,0)$, $(1,0)$. 

$n_{W,\OO}$ is an effective measure of the number of unit cells inside $W$. In general $n_{W,\OO}$ is a sum of the integer number of full unit cells and the irregular, fractional unit cells at the boundaries and defect cores. As explained in \cite{zhang2022pol}, $n_{W,\OO}$ is calculated by fitting the charge response Eq. \ref{QwEq} to the case of atomic insulators, where the individual terms can be computed classically using the fact that the state is described by filled localized Wannier orbitals. This gives the equation
\begin{align}
\label{eq:nwo}
    n_{W,\OO} = r + \vec{L}_{\OO} \cdot \vec{n}_{\OO} + \frac{2\pi - \Gamma}{2\pi} m_{\OO},
\end{align}
for integer $r$. Here, $\vec{n}_{\OO}$ characterizes the fractional unit cells at the edges and dislocations, while $m_{\OO}$ characterizes the fractional unit cells at the corners and disclinations. Their values are tabulated in Table ~\ref{tab:nuc}; see also \cite{zhang2024pol}.

Without excess flux, the total flux $\Phi_W$ in $W$ should be proportional to the area (number of unit cells) $n_{W,\OO}$ and flux per unit cell $\phi$, therefore the excess flux 
$\delta \Phi_{W,\OO}$ is defined as their difference
\begin{align}\label{eq:flux_condition}
\delta\Phi_{W,\OO}\equiv \Phi_W-\phi n_{W,\OO}.
\end{align}

\section*{Model FCI wave functions}\label{sec:parton}

A general way to obtain model wave functions and effective field theories for topologically ordered states is through the parton construction \cite{baskaran1988gauge,jain1989incompressible,wen1991prl,wen1992theory,wen1999projective,lee2006doping,wen2002quantum,barkeshli2010effective,mcgreevy2011fractional,wen04,barkeshli2015particle,sachdev2023quantum}. For example, a $1/k$-Laughlin state $\Psi_{1/k}$ can be obtained by projecting $k$ free fermion many-body wave functions, $\psi_p(\{\vec{r}_i\})$ for $p = 1,\cdots,k$, each describing a Chern insulator with Chern number $C_p= 1$: $\Psi_{1/k}({\{\vec{r}_i\}})= \prod_{p=1}^k \psi_p({\{\vec{r}_i\}})$. Various other non-Abelian, Jain, or hierarchy states can be obtained by choosing other Chern numbers or even more complicated parton constructions \cite{wen1991,Wen1995,wen1999,barkeshli2010effective,barkeshli2015particle,balram2018parton,balram2019parton}. 

\begin{figure*}[t]
    \centering
    \includegraphics[width=17.5cm]{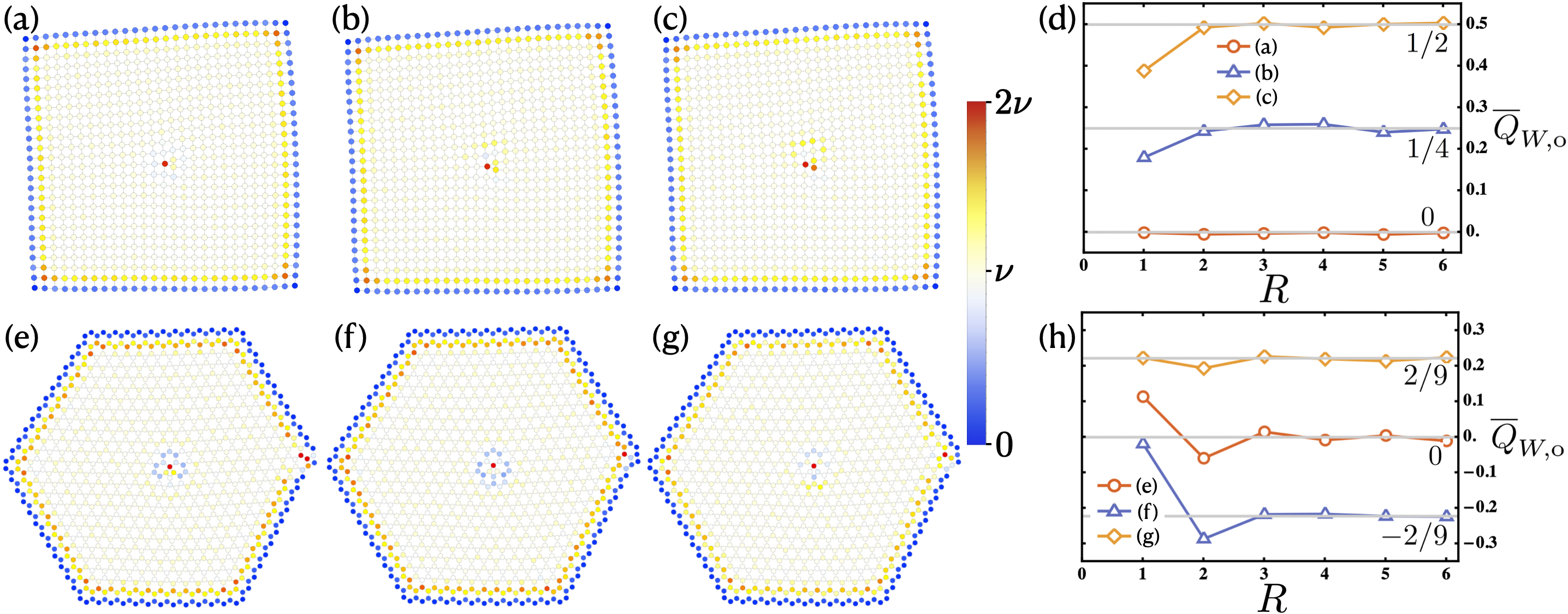}
    \caption{\textbf{(a$\sim$d)}1/2-Laughlin state on a $\vec{b} = (0,1)$ dislocation, filling 90 fermions in the whole disk. \textbf{(a)} Charge profile in units of $\nu$ for $\OO=\alpha$, $\vec{\mathscr{P}}_{\alpha,1}=\vec{\mathscr{P}}_{\alpha,2}=(0,0)$; \textbf{(b)} $\OO=\beta$, $\vec{\mathscr{P}}_{\beta,1}= (0,0), \vec{\mathscr{P}}_{\beta,2}=(1/2,1/2)$; \textbf{(c)} $\OO=\beta$, $\vec{\mathscr{P}}_{\beta,1} = \vec{\mathscr{P}}_{\beta,2} =(1/2,1/2)$. \textbf{(d)} $\overline{Q}_{W,\OO}$ as a function of radius of $W$.
    \textbf{(e$\sim$h)} 1/3-Laughlin state on a $\vec{b} = (1,0)$ honeycomb dislocation, filling 85 fermions in the whole disk. \textbf{(e)} Charge profile in units of $\nu$ for $\OO=\beta_2$, $\vec{\mathscr{P}}_{\beta_2,1}=\vec{\mathscr{P}}_{\beta_2,2}=\vec{\mathscr{P}}_{\beta_2,3}=(0,0)$; \textbf{(f)} $\OO=\beta_2$, $\vec{\mathscr{P}}_{\beta_2,1}= \vec{\mathscr{P}}_{\beta_2,2}=(0,0), \vec{\mathscr{P}}_{\beta_2,3}=(2/3,1/3)$; \textbf{(g)} $\OO=\beta_1$, $\vec{\mathscr{P}}_{\beta_1,1}= \vec{\mathscr{P}}_{\beta_1,2}=(0,0), \vec{\mathscr{P}}_{\beta_1,3}=(1/3, 2/3)$. \textbf{(h)} $\overline{Q}_{W,\OO}$ as a function of radius of $W$.}
    \label{fig:dislocationCom}
\end{figure*}

In the presence of crystalline symmetry, each parton state is labeled by a set of many-body crystalline invariants. These invariants were recently classified systematically for the five orientation-preserving wallpaper groups in \cite{zhang2023complete,manjunath2023classif}. For this paper, we need only keep track of the electric polarization and discrete shift for each parton state, $\vec{\mathscr{P}}_{\OO, p}$, $\mathscr{S}_{\OO,p}$, in addition to the filling (charge per unit cell) $\nu_p$ for each parton state, which must be the same as the filling $\nu$ of the projective wave function: $\nu = \nu_p$. Each parton state has a flux per unit cell $\phi_p$. 
 The projected state effectively has a flux per unit cell $\phi = \sum_p \phi_p$, because the microscopic degrees of freedom are a composite of all $k$ partons. That is, the particle annihilation operator is $c = \prod_{p=1}^k f_i$, where $f_i$ are the individual parton annihilation operators. 

The parton construction introduces a $U(1)^{k-1}$ gauge symmetry, associated with the transformations $f_i \rightarrow e^{i\theta_i}f_i$ which keep the microscopic physical operator $c$ gauge invariant. One can then derive an effective field theory associated with the projected wave function by starting with the effective field theories of each individual parton state and coupling them to a $U(1)^{k-1}$ gauge field. This allows us to determine  $\vec{\mathscr{P}}_{\OO}$ and $\mathscr{S}_{\OO}$ of the FCI in terms of $\vec{\mathscr{P}}_{\OO, p}$, $\mathscr{S}_{\OO,p}$ for each parton state. We carry out this calculation in Appendix \ref{app:partonMF}, with the result:
\begin{align}
\label{partonSP}
\mathscr{S}_{\OO} &= \frac{1}{k} \sum_{p=1}^m \mathscr{S}_{\OO, p} \mod  M_\OO/k, 
\nonumber \\
\vec{\mathscr{P}}_{\OO}&= \frac{1}{k} \sum_{p=1}^m \vec{\mathscr{P}}_{\OO, p} \mod \frac{1}{k} \mathbb{Z}^2
\end{align}
Ref. \cite{zhang2022pol} showed how $\vec{\mathscr{P}}_{\OO, p}$ and $\mathscr{S}_{\OO, p}$ transform linearly under a shift of origin $\OO \rightarrow \OO + \vec{v}$. Linearity of Eq. \ref{partonSP} then implies the same transformation rules apply to $\vec{\mathscr{P}}_{\OO}$ and $\mathscr{S}_{\OO}$ for the projected states. These transformation rules are summarized in Table \ref{table:shiftingO} in Appendix~\ref{app:reviewTorsion} .

\section*{Numerical Calculations}\label{sec:num}

We now numerically verify the charge response Eq. ~\eqref{QwEq} and demonstrate the universal fractionally quantized contributions from lattice disclinations, dislocations, corners, and boundaries. We use model wave functions obtained using the parton construction. Depending on the parton invariants, $\mathscr{S}_{\OO,p}$ and $\vec{\mathscr{P}}_{\OO,p}$, we can describe $1/k$-Laughlin states with different values of quantized electric polarization $\vec{\mathscr{P}}_{\OO}$ and discrete shift $\mathscr{S}_{\OO}$ to validate the general theory. 
In this section, we will focus on 
1/2-Laughlin states on the square lattice with disclinations, dislocations, edges or corners. We set $\OO=\alpha,\beta$ where $M_{\OO}=4$. We also study $1/3$-Laughlin states on the honeycomb lattice with disclinations and dislocations for origins $\OO=\beta_1,\beta_2$ where $M_{\OO}=3$, and $\OO=\alpha$ where $M_{\alpha}=6$.

For the parton states, we use the ground states of the Harper-Hofstadter model of free fermions with flux $\phi_p$ per unit cell and filling $\nu_p$. See Appendix \ref{app:num_method} for additional details. For the square lattice with $M = 4$, we use the convention that $\beta$ represents vertices and $\alpha$ represents plaquette centers. For the honeycomb lattice with $M = 6$, the only choice of unit cell places $\beta_1$ and $\beta_2$ at the vertices, $\alpha$ at the plaquette centers (see Fig. \ref{fig:unit_cells}).

To obtain the FCI wave functions on the lattices with defects, we first construct the free fermion parton states on the lattices with defects, using methods discussed in \cite{zhang2022fractional,zhang2022pol}, and then apply the projection. 

In each set of calculations, we sample $5\times 10^7$ ground state configurations using the Metropolis Monte Carlo method, as reviewed in Appendix \ref{app:num_method}. Additional numerical results are presented in Appendix \ref{app:num}.

{\bf Disclination charge.}
As shown in Fig.~\ref{fig:disclinationCom}(a$\sim$c), we place the 1/2-Laughlin state on a square lattice with a $\Gamma=\pi/2$ disclination. We consider vertex-centered disclinations (Fig.~\ref{fig:disclinationCom}(a)) with $\OO = \beta$, Burgers vectors $\vec{b}_{\beta} \simeq (0,0)$ and plaquette-centered disclinations (Figs ~\ref{fig:disclinationCom}(b),(c)) with $\OO = \alpha$, $\vec{b}_{\alpha} \simeq (0,0)$. This allows us to extract the invariant $\mathscr{S}_{\OO}$ from the charge response.

We define the parton Hamiltonian such that $\delta\Phi_{W,\OO}=0$, eliminating the $\sigma_{H}\delta \Phi_{W,\OO}$ contribution to the charge response. To achieve this, for $\OO=\beta$, we demand that each parton experiences a $\phi_p$ flux per plaquette. For $\OO=\alpha$, we demand that each parton have $\phi_p$ per regular plaquette and $\frac{3}{4}\phi_p$ flux through the central triangular plaquette, according to Eq.~\eqref{eq:nwo}. This choice ensures that $\delta\Phi_{W,\OO}=\Phi_W-\phi n_{W,\OO}=0$.

We define the regularized charge $\overline{Q}_{W,\OO}$ as:

\begin{align}
    \overline{Q}_{W,\OO}=Q_W-\nu n_{W,\OO}.
\end{align}
This removes the background contribution from the charge so that $\overline{Q}_{W,\OO}$ converges to certain fractions for sufficiently large $W$. Based on Eq.~\eqref{QwEq}, when $\vec{b}_{\OO} \simeq (0,0$), $\overline{Q}_{W,\OO}$ should converge to

\begin{align}
    \overline{Q}_{W,\OO}=\mathscr{S}_{\OO}/4 \mod 1/2,
\end{align}
which is consistent with the numerical result (Fig.~\ref{fig:disclinationCom}(d)).

\begin{figure*}[t]
    \centering
    \includegraphics[width=17.5cm]{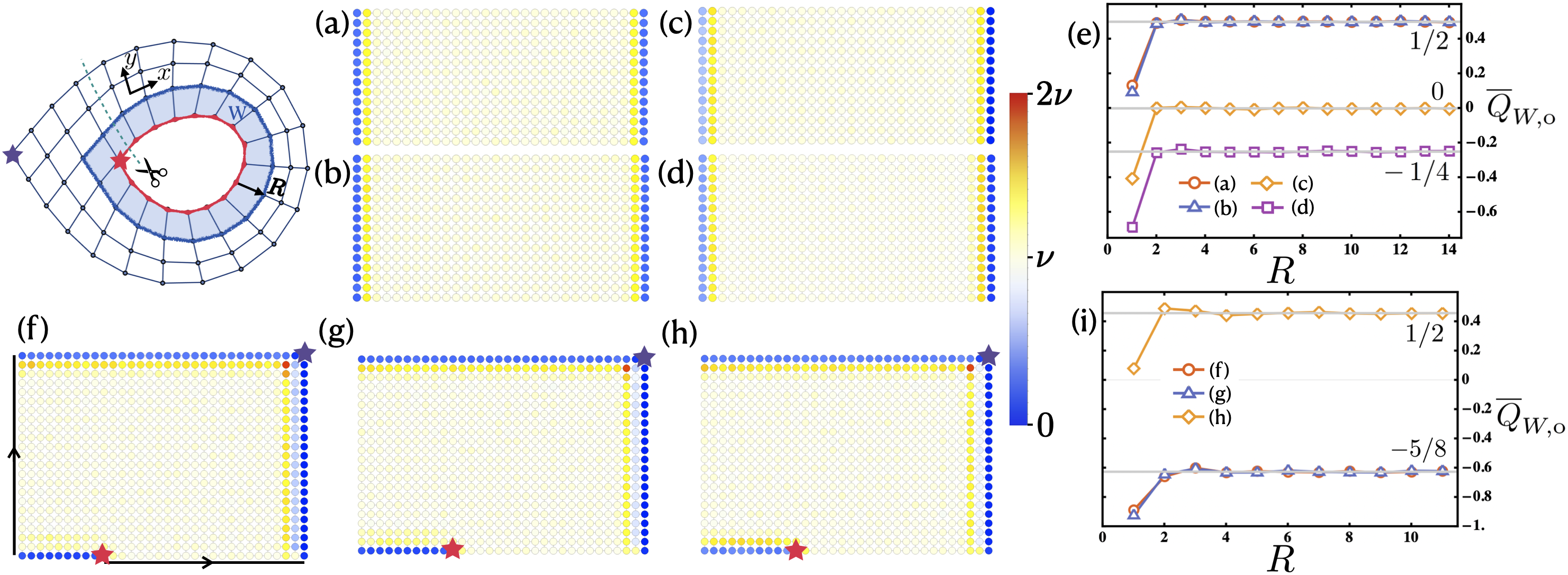}
    \caption{\textbf{(a$\sim$e)}1/2-Laughlin state on cylinders, filling 60 fermions in the whole disk. We calculate $Q_W$ where $W$ covers the left boundary. \textbf{(a)} The charge profile in units of $\nu$ with periodic boundary condition in the $y-$direction for $\OO=\alpha$, $\vec{L}_{\alpha}=(0,14)$, $\vec{\mathscr{P}}_{\alpha,1}=\vec{\mathscr{P}}_{\alpha,2}=(0,0)$; \textbf{(b)} same as (a) but with $\vec{L}_{\alpha}=(0,15)$; \textbf{(c)} $\OO=\beta$, $\vec{L}_{\beta}=(0,14)$,  $\vec{\mathscr{P}}_{\beta,1} = (1/2,1/2)$, $\vec{\mathscr{P}}_{\beta,2} =(0,0)$; \textbf{(d)} same as (c) but with $\vec{L_\OO}=(0,15)$
    \textbf{(e)} $\overline{Q}_{W,\OO}$ as a function of radius of $W$.
    \textbf{(f$\sim$i)}1/2-Laughlin state on a ribbon geometry, filling 80 fermions in the whole disk. $W$ covers the inner corner of the ribbon. {(f)} Charge profile in units of $\nu$. We glue the two boundaries along the arrow direction to obtain the ribbon geometry in the top left, where the inner and outer corners are labeled with stars. Here $\OO=\beta$, $\Gamma=\frac{2\pi}{4}$, $\vec{L}_{\OO}\simeq(0,0)$, $\vec{\mathscr{P}}_{\beta,1}=\vec{\mathscr{P}}_{\beta,2}=(0,0)$, $\mathscr{S}_{\beta,1}=\mathscr{S}_{\beta,2}=1/2$; \textbf{(g)} same as (f) but with $\vec{L}_{\OO}\simeq(0,1)$. \textbf{(h)} $\OO=\alpha$, $\vec{L}_{\OO}\simeq(0,0)$, $\vec{\mathscr{P}}_{\alpha,1}=\vec{\mathscr{P}}_{\alpha,2}=(0,0)$, $\mathscr{S}_{\alpha,1}=1/2, \mathscr{S}_{\alpha,2}=7/2$. \textbf{(i)} $\overline{Q}_{W,\OO}$ as a function of radius of $W$.}
    \label{fig:edgeCorner12}
\end{figure*}

As shown in Fig. \ref{fig:disclinationCom}(e$\sim$g), we place the 1/3-Laughlin state on a honeycomb lattice with a $\Gamma=2\pi/3$ disclination. We consider vertex-centered disclinations (Fig. \ref{fig:disclinationCom}(e)) with $\OO = \beta_1$, Burgers vector $\vec{b}_{\beta_1} \simeq (0,0)$ and plaquette-centered disclinations (Fig. \ref{fig:disclinationCom}(f)(g)) with $\OO = \alpha$, $\vec{b}_\alpha \simeq (0,0)$. 
To ensure $\delta{\Phi_{W,\OO}}=0$, we require that for each parton state the irregular plaquette at the center of the defect for $\OO=\alpha$ has $\frac{2}{3}\phi_p$ through it. Again, when $\vec{b}_\OO \simeq (0,0)$, the regularized charge $\overline{Q}_{W,\OO}$ is predicted to converge to
\begin{equation}
    \overline{Q}_{W,\OO}=\mathscr{S}_{\OO}/3 \mod 1/3,
\end{equation}
which aligns with the numerical result.

Note that $\overline{Q}_{W,\OO}$ can change trivially by the minimal anyon charge, which is either $1/2$ or $1/3$ in these two examples. Numerically, this is achieved by filling a trivial state with charge 1 at the disclination core for one of the three parton states. This is consistent with the fact that $\mathscr{S}_\OO$ is only defined modulo $M_\OO/m$.

{\bf Dislocation charge.} 
As depicted in Fig.~\ref{fig:dislocationCom}(a$\sim$c) we place the $1/2$-Laughlin state on a square lattice with a dislocation with $\vec{b}_{\OO}=(0,1)$ to extract the topological invariant $\PO$ from the charge response. We again define the parton Hamiltonian such that $\delta\Phi_{W,\OO}=0$. This means the flux through the central triangular plaquette is set to be $0$ if $\OO=\alpha$, and $\phi_p/2$ if $\OO=\beta$ following Eq.~\eqref{eq:nwo}. Based on Eq.~\eqref{QwEq}, the regularized charge $ \overline{Q}_{W,\OO}$ should converge to

\begin{align}
    \overline{Q}_{W,\OO}=\mathscr{P}_{\OO,y} \mod 1/2,
\end{align}
which agrees with the numerical results in Fig.~\ref{fig:dislocationCom}(d). 

As shown in Fig.~\ref{fig:dislocationCom}(e$\sim$g), we place the 1/3-Laughlin state on a honeycomb lattice with a dislocation with $\vec{b}_{\OO}=(1,0)$. To ensure $\delta\Phi_{W,\OO}=0$, we require that the central irregular square plaquette has $\phi_p/3$ flux if $\OO=\beta_1$, and $2\phi_p/3$ flux if $\OO=\beta_2$. $\overline{Q}_{W,\OO}$ should then converge to

\begin{align}
    \overline{Q}_{W,\OO}=\mathscr{P}_{\OO,x} \mod 1/3,
\end{align}
which again agrees with the numerical results.

{\bf Boundary and corner charge.}
Finally, we test the charge response at edges and corners of the lattice for the 1/2-Laughlin state.
As shown in Fig.~\ref{fig:edgeCorner12}(a$\sim$d), we place the 1/2-Laughlin state on cylinders with different $\vec{L}_{\OO}=(0,L_{\OO,y})$, and calculate the $\overline{Q}_{W,\OO}$ at the left boundary of the cylinder.  To enforce $\delta\Phi_{W,\OO}=0$, we set the holonomy for the parton states along the left boundary of the lattice to 
$0$ for $\OO=\alpha$ and $\phi_p L_{\OO,y}/2$ for $\OO=\beta$ according to Eq. \eqref{eq:nwo}. The regularized charge $\overline{Q}_{W,\OO}$ should converge to
\begin{align}\label{eq:edgeCharge}
    \overline{Q}_{W,\OO}=\SO+L_{\OO,y}\mathscr{P}_{\OO,y} \mod 1/2,
\end{align}
matching the numerical results. Note $\SO \mod 1/2 = 0$.

The above result is robust to random perturbations that break the translation symmetry along the boundary, since such perturbations cannot change the total charge on the boundary mod $1/2$. A similar robustness was demonstrated for integer Chern insulators in \cite{zhang2024pol}.

Next, we demonstrate how to extract $\mathscr{S}_{\OO}$ from the corner charge response. To isolate a corner contribution, we want only one corner within the region $W$. Furthermore, because of the chiral edge states, the charge will be delocalized along the edge, so we must define $W$ in such a way that its boundary does not intersect with the boundary of the lattice. 
We thus consider the 1/2-Laughlin state on a ribbon geometry, where $W$ covers the inner corner with $\Gamma = -\pi/2$, as shown in the top left of Fig.~\ref{fig:edgeCorner12}. We enforce $\delta\Phi_{W,\OO}=0$ by setting the holonomy for the parton state in the inner loop (colored red) to be
$\phi_pn_{W,\OO}$ where $n_{W,\OO}$ is as defined in Eq.~\eqref{eq:nwo}. Based on Eq.~\eqref{QwEq}, the regularized charge $ \overline{Q}_{W,\OO}$ should converge to
\begin{align}\label{eq:ribbonCharge}
    \overline{Q}_{W,\OO}= \SO - \frac{\SO}{4} + \vec{L}_{\OO}\cdot \vec{\mathscr{P}}_{\OO} \mod 1/2,
\end{align}
which agrees with the numerical results in Fig.~\ref{fig:edgeCorner12}(f$\sim$i). 

\section*{Discussion}

Our results demonstrate how microscopic charge measurements can be used to extract the invariants $\vec{\mathscr{P}}_{\OO}$ and $\mathscr{S}_{\OO}$. These invariants were originally predicted in \cite{manjunath2020FQH,manjunath2021cgt} using the more abstract setting of topological quantum field theory and $G$-crossed modular category theory \cite{barkeshli2019}; the agreement of these abstract methods with concrete microscopic computations is remarkable. Note that these invariants can also be extracted in FCIs using partial rotation operations \cite{kobayashi2024crystalline}, giving a remarkable example of duality in quantum many-body systems.  The fractional quantized electric polarization $\vec{\mathscr{P}}_{\OO}$ of FCIs, along with the o-dependence of $\mathscr{S}_\OO$, has no analog in continuum FQH states \cite{Wen1992shift,manjunath2020FQH} and are thus invariants fundamentally tied to the crystalline setting. 

\section*{Acknowledgements}

We thank Naren Manjunath for discussions and collaboration on related work. This work is supported by NSF DMR-2345644 and NSF
QLCI grant OMA-2120757 through the Institute for Robust Quantum Simulation.

\appendix

\section{Review of discrete torsion vector, spin vector, polarization, and discrete shift}\label{app:reviewTorsion}

In this section, we provide a brief summary of the spin vector, discrete torsion vector, their various equivalence relations, and how they relate to the polarization and discrete shift. In Appendix \ref{app:response}, we will explain the origin of these topological invariants in terms of the topological field theory. 

Consider an Abelian topological order given by a $D \times D$ $K$-matrix. The symmetry fractionalization is partially specified by a $D$-component integer charge vector $\vec{v}$, spin vector $\vec{s}_{\OO}$ and a $2D$-component discrete torsion vector $\vec{t}_{\OO} = (\vec{t}_{\OO, x}, \vec{t}_{\OO,y})$. In addition to symmetry fractionalization, we can consider additional SPT contributions (see Eq. \ref{topaction}). The SPT will be characterized by $\mathscr{S}_{SPT, \OO} \mod M$ and $\vec{\mathscr{P}}_{SPT, \OO} \mod \mathbb{Z}^2$, among other invariants. In this section and the next, we will always assume that $M_{\OO} = M$ (that the order of the point group at $\OO$ is $M$).

We thus have a tuple $(\mathscr{S}_{SPT,\OO}, \vec{\mathscr{P}}_{SPT, \OO}, \vec{s}_\OO, \vec{t}_{\OO})$. As we explain in Appendix \ref{app:response}, this tuple is subject to equivalence relations:
\begin{align}
    (\mathscr{S}_{SPT,\OO}, \vec{s}_\OO) &\sim 
    (\mathscr{S}_{SPT,\OO} - \vec{v}^T \vec{\Lambda}, \vec{s}_\OO + K \vec{\Lambda}) 
    \nonumber \\
    &\sim (\mathscr{S}_{SPT,\OO}, \vec{s}_\OO + M \vec{\Lambda}) 
\end{align}
The discrete shift is given by
\begin{align}
    \mathscr{S}_{\OO} = \mathscr{S}_{SPT, \OO} + \vec{v}^T K^{-1} \vec{s}_{\OO}.
\end{align}
The above equivalences imply
\begin{align}
    \mathscr{S}_{\OO} \sim \mathscr{S}_{\OO} + M \vec{v}^T K^{-1} \vec{\Lambda} \sim  \mathscr{S}_{\OO}  + M,
\end{align}
meaning $\mathscr{S}_{\OO}$ is defined modulo $M$.
For the $1/2$-Laughlin state with $M = 4$, we take $K = 2$, $v =1$, and the invariant is $\mathscr{S}_{\OO} \mod 2$, and there are four possible values, $0, 1/2, 1, 3/2 \mod 2$. 

For bosonic theories, $K$ is even on the diagonals, and $\vec{s}_{\OO}$ is an integer vector. For fermionic theories, where $K$ can have diagonal entries that are odd, we have an additional requirement, which is that the physical fermion should have a half-integer orbital spin.\footnote{This can be understood from the following perspective. In the topological field theory, each fermion already has a half-integer topological spin. Therefore a $2\pi$ rotation would imply a $(-1)$ phase on the fermions. However, if our microscopic symmetry operator satisfies $C_M^M = +1$, the fermions need to carry an additional orbital angular momentum of $1/2 \mod 1$ so that the $2\pi$ rotation overall gives a $+1$ phase on the physical fermion.} This translates to the requirement that if $\vec{\Lambda}^T K \vec{\Lambda}$ is odd, then $\vec{s}_{\OO} \cdot \vec{\Lambda} = 1/2 \mod 1$, for any integer vector $\vec{\Lambda}$. Therefore, for the $1/3$-Laughlin state with $M = 3$, we take $K = 3$, $v = 1$, and $s_{\OO}$ to be half-integer. The discrete shift is then $\mathscr{S}_{\OO} = \mathscr{S}_{SPT,\OO} + s_\OO/3 \mod 1 = s_{\OO}/3 \mod 1$. There are thus 3 possible inequivalent values, $\mathscr{S}_{\OO} = 1/6, 1/2, 5/6 \mod 1$. 

\begin{table}[]
\def\arraystretch{1.1}
    \centering
    \begin{tabular}{c||c|c|c|c}
    \hline
    $M$ & 2& 3& 4&6 \\ \hline
     $U(2\pi/M)$ & $\begin{pmatrix}-1 & 0 \\ 0 & -1\end{pmatrix}$ & $\begin{pmatrix}-1 & -1 \\ 1 & 0\end{pmatrix}$ &  $\begin{pmatrix}0 & -1 \\ 1 & 0\end{pmatrix}$ & $\begin{pmatrix}0 & -1 \\ 1 & 1\end{pmatrix}$  \\ \hline
    \end{tabular}
    \caption{Elementary rotation matrices $U(2\pi/M)$, for the coordinate basis shown in Fig.~\ref{fig:unit_cells}.}
    \label{tab:U_mats}
\end{table}

\bgroup
\def\arraystretch{1.8}
\begin{table}
    \centering
\begin{tabular}{ c||l|l  }

\hline

$M$ &
$\mathscr{S}_{\text{o}+\vec{v}}$ &  $\vec{\mathscr{P}}_{\text{o}+\vec{v}}$ \\
\hline
2& $
   \mathscr{S}_{\text{o}} - 4 \vec{v}\cdot\vec{\mathscr{P}}_{\text{o}} + 4\kappa(v_x^2 + v_y^2 + v_x v_y)
$&$\vec{\mathscr{P}}_{\text{o}}+
(-v_y \kappa,  v_x \kappa)$
\\
\hline
4 & $
   \mathscr{S}_{\text{o}} +  4\mathscr{P}_{\text{o},x} - \kappa
$ & $\vec{\mathscr{P}}_{\text{o}}+(-\frac{1}{2} \kappa,  \frac{1}{2} \kappa)$\\

\hline
3 & $\mathscr{S}_{\text{o}}- 9 v_y \mathscr{P}_{\text{o},x} -3\kappa (v_x^2 + v_y^2 + v_x v_y)$ & $\vec{\mathscr{P}}_{\text{o}}+(-v_y \kappa,  v_x \kappa)$\\
\hline
6 & $\mathscr{S}_{\text{o}}$ & 0\\
\hline

\end{tabular}
\caption{Transformation of $\mathscr{S}_\text{o}$ and $\vec{\mathscr{P}}_{\text{o}}$ under $\text{o}\rightarrow \text{o}+(v_x,v_y)$, which shifts the origin from one $C_M$ symmetric point to another $C_M$ symmetric point. Note that $\mathscr{S}_{\text{o}}$ and $\vec{\mathscr{P}}_{\text{o}}$ are only defined up to equivalences as described in the main text. For $M=4$ we have taken the unique non-trivial choice $\vec{v}=(\frac{1}{2},\frac{1}{2}).$
}\label{table:shiftingO}

\end{table}
\egroup

As we explain in Appendix \ref{app:response}, the discrete torsion vector $\vec{t}_{\OO}$ has an equivalence relation 
\begin{align}
(\vec{\mathscr{P}}_{SPT,\OO},\vec{t}_{\OO,I}) &\sim  (\vec{\mathscr{P}}_{SPT,\OO},\vec{t}_{\OO,I} + (1 - U(2\pi/M)^T) \vec{\Lambda}), \nonumber\\
&\hspace{3cm}
\nonumber \\
(\vec{\mathscr{P}}_{SPT,\OO},\vec{t}_{\OO,i}) &\sim (\vec{\mathscr{P}}_{SPT,\OO}- \vec{v}^T \vec{\Lambda},\vec{t}_{\OO,i} + K \vec{\Lambda}) , \\
&\hspace{3cm} \nonumber
\end{align}
for $I = 1,\cdots, D$ and $i = x,y$. The rotation matrices are tabulated in Table~\ref{tab:U_mats}.\footnote{Note that compared to \cite{manjunath2021cgt}, our formulas are in terms of $U(2\pi/M)^T$ instead of $U(2\pi/M)$, which originates from different conventions. }
The electric polarization is given by 
\begin{align}
    \vec{\mathscr{P}}_{\OO,i}=[1-U(2\pi/M)^T]^{-1}_{ij }\vec{v}^TK^{-1}\vec{t}_{\OO,j}+\vec{\mathscr{P}}_{SPT,\OO},
\end{align}
for $i\in x,y$, and repeated indices are summed implicitly. From the above equivalences, we get
\begin{align}
    \PO\sim \PO+\vec{v}^TK^{-1} \vec{\Lambda}\sim \PO+\vec{\Lambda}.
\end{align}
For the case of the $1/k$-Laughlin state with Hall conductivity $\sigma_H = \frac{1}{k} \frac{1}{2\pi}$, $K = k$ and $v = 1$ so that the equivalence relation simplifies to $\vec{\mathscr{P}}_{\OO} \mod (1/k,1/k)$, which we also denote as $\vec{\mathscr{P}}_{\OO} \mod  \mathbb{Z}^2$.

For the $1/2$-Laughlin state on a square lattice with $\sigma_H = \frac{1}{2} \frac{1}{2\pi}$, we have
\begin{align}
    \nonumber\PO&= [1-U(2\pi/4)^T]^{-1}  \frac{1}{2}\vec{t}_{\OO}+\vec{\mathscr{P}}_{SPT,\OO} \mod \frac{1}{2}\Z^2\\
    &=  [1-U(2\pi/4)^T]^{-1} \frac{1}{2}\vec{t}_{\OO}\mod \frac{1}{2}\Z^2 ,
\end{align}
Therefore the inequivalent values are $\PO=(0,0)$, $(1/4,1/4) \mod \frac{1}{2} \mathbb{Z}^2$. 

For a 1/3-Laughlin state on a honeycomb lattice with $\sigma_H = \frac{1}{3} \frac{1}{2\pi}$, 
\begin{align}
    \nonumber\PO&= [1-U(2\pi/3)^T]^{-1} \frac{1}{3}\vec{t}_{\OO}+\vec{\mathscr{P}}_{SPT,\OO} \mod \frac{1}{3}\Z^2\\
    &=  [1-U(2\pi/3)^T]^{-1}  \frac{1}{3}\vec{t}_{\OO} \mod \frac{1}{3}\Z^2,
\end{align}
The inequivalent values are $\PO = (0,0)$, $ (1/9,2/9)$, $(2/9,1/9) \mod \frac{1}{3} \mathbb{Z}^2$.

The dependence of $\mathscr{S}_\OO$ and $\vec{\mathscr{P}}_\OO$ on $\OO$, shown in Table \ref{table:shiftingO}, was presented in \cite{zhang2022pol} for invertible states on the basis of both empirical numerical results and $G$-crossed braided tensor category computations. In Appendix \ref{app:partonMF}, we will see that these transformation rules continue to hold by linearity for FCI states described by the parton construction.  

\hspace{3cm}
\section{Review of topological crystalline gauge theory}\label{app:response}

In this section, we review the topological crystalline gauge theory derived in \cite{manjunath2021cgt} to describe crystalline invariants for Abelian topological orders. We consider Abelian topological orders, which can be described by a $U(1)^D$ Chern-Simons theory and an integer symmetric $K$-matrix. For now we consider bosonic topological orders, for which the diagonal entries of $K$ are even integers. We discuss the fermionic case afterwards. 

We consider the global symmetry group
\begin{equation}
    G = U(1) \times_\phi [\mathbb{Z}^2 \rtimes \mathbb{Z}_M],
\end{equation}
where $\mathbb{Z}_M$ (for $M = 2, 3, 4, 6$) represents the point group rotational symmetry about a fixed point $\OO$. The $\mathbb{Z}^2$ denotes translation symmetry in the $x$ and $y$ directions, generated by $\tilde{T}_x$ and $\tilde{T}_y$. Here, $\phi$ is the $U(1)$ flux per unit cell. The notation $\times_\phi$ indicates a central extension of $\mathbb{Z}^2$ by $U(1)$, corresponding to the magnetic translation algebra: $\tilde{T}_y^{-1}\tilde{T}_x^{-1}\tilde{T}_y\tilde{T}_x=e^{i\phi\hat{N}}$, where $\hat{N}$ is the number operator.  

As discussed in \cite{manjunath2021cgt,manjunath2020FQH}, topological phases with crystalline symmetry can be characterized and classified by developing a topological field theory with a background gauge field for $G$. Specifically, we have a $G$ gauge field $B = (A, \vec{R}, \omega_\OO)$, where $A$ is a background $U(1)$ gauge field, $\vec{R}$ is a background $\mathbb{Z}^2$ gauge field, and $\omega_\OO$ is a background $\mathbb{Z}_M$ gauge field for the $M$-fold rotations about $\OO$. In what follows, for ease of notation we will drop the subscript $\OO$ in $\omega_\OO$, but keep it in the coupling constants. 

To develop the topological field theory, we first triangulate the space-time manifold $\mathcal{M}^3$. Then we define gauge fields on 1-simplices $ij$ of the triangulation. That is, we have $A_{ij} \in \mathbb{R}$, $\vec{R}_{ij} \in 2\pi \mathbb{Z}^2$, $\omega_{ij} \in \frac{2\pi}{M} \mathbb{Z}$. Technically $A$ and $\omega$ are lifts of $U(1)$ and $\mathbb{Z}_M$ to $\mathbb{R}$ and $\frac{2\pi}{M} \mathbb{Z}$, respectively. 

The topological terms can be derived in the following way. We start by considering an invertible theory with $U(1)^D \times G$ symmetry. The topological terms are classified by group cohomology:\cite{dijkgraaf1990,Chen2013}
\begin{widetext}
\begin{align}
\mathcal{H}^3(U(1)^D \times G, \mathbb{R}/\mathbb{Z}) \simeq \mathcal{H}^3(U(1)^D, \mathbb{R}/\mathbb{Z}) \oplus \mathcal{H}^2(G, \mathbb{Z}^D) \oplus \mathcal{H}^3(G, \mathbb{R}/\mathbb{Z}). 
\end{align}
\end{widetext}
Here we have used the K\"unneth decomposition 
\begin{align}
\mathcal{H}^3(U(1)^D \times G, \mathbb{R}/\mathbb{Z}) = \bigoplus_{k=0}^3 \mathcal{H}^{3-k}(G, \mathcal{H}^k(U(1)^D, \mathbb{R}/\mathbb{Z})), 
\end{align}
together with the fact that $\mathcal{H}^1(U(1)^D, \mathbb{R}/\mathbb{Z}) = \mathbb{Z}^D$ and $\mathcal{H}^2(U(1)^D, \mathbb{R}/\mathbb{Z})$ is trivial. An element of $\mathcal{H}^3(U(1)^D \times G, \mathbb{R}/\mathbb{Z})$ can thus be written as $([\alpha_K], [\alpha_\text{frac}], [\alpha_{\text{SPT}}])$, with $[\alpha_K] \in \mathcal{H}^3(U(1)^D, \mathbb{R}/\mathbb{Z})$, $[\alpha_{\text{frac}}] \in  \mathcal{H}^2(G, \mathbb{Z}^D)$, and $[\alpha_{\text{SPT}}] \in \mathcal{H}^3(G, \mathbb{R}/\mathbb{Z})$. 

Given flat $U(1)^D$ gauge fields $a^I$, $I = 1,\cdots, D$, and a flat $G$ gauge field $(A, \vec{R}, \omega)$, we can pull back these cohomology classes to the space-time manifold to get a topological action with Lagrangian density 
\begin{widetext}
\begin{align}
\label{topaction}
    \mathcal{L} &= -\frac{1}{4\pi} K_{IJ} a^I \cup d a^J + \mathcal{L}_{\text{frac}} + \mathcal{L}_{SPT} ,
    \nonumber \\
    \mathcal{L}_{\text{frac}} &= \frac{1}{2\pi} a^I \cup (v_I d A + s_{\OO,I} d \omega + \vec{t}_{\OO,I} \cdot (1 - U(2\pi/M)^T)^{-1} d \vec{R} + m_I A_{XY}) ,
    \nonumber \\
    \mathcal{L}_{SPT} &= \frac{\bar{\sigma}_{H,SPT}}{4\pi}A \cup dA+\frac{\mathscr{S}_{SPT, \OO}}{2\pi}A \cup d\omega + 
    \frac{\vec{\mathscr{P}}_{SPT,\OO}}{2\pi} \cdot A \cup d \vec{R}
    + \frac{\kappa_{SPT}}{2\pi}A \cup A_{XY} + \cdots    
\end{align}
\end{widetext}
Here $d$ is the coboundary operator, $\cup$ denotes the cup product, and $A_{XY}$ is a 2-cochain whose explicit formula in terms of $\vec{R}$ and $\omega$ is given in \cite{manjunath2021cgt}. The $\cdots$ refers to additional terms that do not affect the charge response and do not concern us in this paper; see \cite{manjunath2021cgt,manjunath2020FQH} for details. Repeated indices are implicitly summed over. The $K$-matrix specifies the cohomology class $[\alpha_K]$. The SPT coefficients $\mathscr{S}_{SPT, \OO}$, $\kappa_{SPT}$ are integers, $\bar{\sigma}_{H,SPT}$ is an even integer, and $(1 - U(2\pi/M)^T) \vec{\mathscr{P}}_{SPT,\OO}$ is an integer vector. Similarly, $\vec{v}$, $\vec{s}_\OO$, $\vec{m}$, $\vec{t}_{\OO, x},\vec{t}_{\OO,y}$ are $D$-component integer vectors. 

Note that since $\omega$ describes a $\mathbb{Z}_M$ gauge field, $M \omega$ is trivial, so we have an equivalence
\begin{align}
    s_{\OO,I} \simeq s_{\OO,I} + M .
\end{align}
In particular, a term of the form $M \int_{\mathcal{M}^3} a^I \cup d \omega = M \int_{\mathcal{M}^3} \omega \cup d a^I = 2\pi \mathbb{Z}$ is trivial on a closed space-time manifold $\mathcal{M}^3$. Analogously, $(1 - U(2\pi/M)^T)\vec{R}$ is also trivial, which implies the equivalence
\begin{align}
    \vec{t}_{\OO,I} \sim \vec{t}_{\OO,I} + (1 - U(2\pi/M)^T)\vec{\Lambda},  
\end{align}
where $\vec{\Lambda}$ is an integer vector.

Next, we gauge the $U(1)^D$ symmetry by promoting $a$ to a dynamical $U(1)^D$ gauge field. The symmetry fractionalization is specified by the terms in $\mathcal{L}_{\text{frac}}$, involving the charge vector $\vec{v}$, the spin vector $\vec{s}_{\OO}$, the discrete torsion vector $\vec{t}_{\OO}$, and the area vector $\vec{m}$, which are all integer vectors. Since $a$ is dynamical, we can shift the gauge fields $a^I \rightarrow a^I + \Lambda_I A$, where $\Lambda_I$ is an integer, without affecting the path integral. This induces the remaining equivalence relations mentioned in Appendix \ref{app:reviewTorsion}, and has the effect of reducing the classification of topological terms in $\mathcal{L}_{\text{frac}}$ to $\mathcal{H}^2(G, \mathbb{Z}^D/K\mathbb{Z}^D) \simeq \mathcal{H}^2(G, \mathcal{A})$, which is the known classification of symmetry fractionalization. Here $\mathcal{A} \simeq \mathbb{Z}^D/K\mathbb{Z}^D$ is the Abelian group formed by fusion of the Abelian anyons. 

Fermionic topological orders, where $K$ can have diagonal entries that are odd, are treated similarly, with a few minor modifications. As described in Appendix \ref{app:reviewTorsion}, the spin vector $\vec{s}_{\OO}$ should have the property that $\vec{s}_{\OO} \cdot \vec{\Lambda} = 1/2 \mod 1$ whenever $\vec{\Lambda}^T K \vec{\Lambda}$ is odd. We also can assume without loss of generality that the SPT contributions correspond to stacking with fermionic SPTs which have zero chiral central charge, in which case all other couplings have the same quantization as in the bosonic case. We note that a more systematic derivation of the topological field theory for fermionic systems, which also applies to non-Abelian topological orders, can use the more sophisticated categorical formalism developed in \cite{bulmashSymmFrac,barkeshli2021invertible,aasen2021characterization}. 

To understand the topological response properties, we formally integrate out the dynamical gauge field $a$ to get a topological response action
\begin{align}
    \mathcal{L}_{\text{resp}} = &\frac{\sigma_H}{2} A \cup d A + \frac{\mathscr{S}_{\OO}}{2\pi} A \cup d \omega + \frac{\vec{\mathscr{P}}_\OO}{2\pi} A \cup d \vec{R}
    \nonumber \\
    & + \kappa A \cup A_{XY} + \cdots,
\end{align}
where
\begin{align}\label{eq:invariants}
    \sigma_H &= \frac{1}{2\pi} (\vec{v}^T K^{-1} \vec{v} + \bar{\sigma}_{H,SPT})
    \nonumber \\
    \mathscr{S}_\OO &= \vec{v}^T K^{-1} \vec{s}_{\OO} + \mathscr{S}_{SPT, \OO} 
    \nonumber \\
    \nonumber  \vec{\mathscr{P}}_{\OO,i}&=[1-U(2\pi/M_{\OO})^T]^{-1}_{ij }\vec{v}^TK^{-1}\vec{t}_{\OO,j}+\vec{\mathscr{P}}_{SPT,\OO}\\
    \kappa&= \vec{v}^T K^{-1} m+ \kappa_{SPT}
\end{align}
The above implies a charge response
\begin{align}\label{eq:calculateResponse}
    Q_W &= \int_W \frac{\delta \mathcal{L}}{\delta A_0} = \sigma_H \frac{\Phi_W}{2\pi} + \mathscr{S}_\OO \frac{\Omega_{\text{disc}}}{2\pi} + \vec{\mathscr{P}}_{\OO} \cdot \vec{b} + \kappa n_{W},
\end{align}
where $\Phi_W = \int_W d A$ is the flux in $W$, $\Omega_{disc} = \int_W d \omega$ is the disclination angle in $W$ (note that $\omega$ was a lift, so $\Omega_{disc}$ is defined as a real number, not just modulo $2\pi$), $\vec{b} = \frac{1}{2\pi}\int_W d \vec{R}$ is the Burgers vector in region $W$,  $n_W = \frac{1}{2\pi}\int_W A_{XY}$ is the number of unit cells in $W$, and $A_0$ is the time-component of $A$. Defining $\delta\Phi_{W} := \Phi_W - \phi n_W$ and using the relationship between the filling $\nu$ and $\kappa$:
\begin{align}
    \nu = \phi \sigma_H + \kappa,
\end{align}
Eq.~\eqref{eq:calculateResponse} can be expressed as
\begin{align}
    Q_W =\sigma_H \frac{\delta\Phi_{W}}{2\pi} + \mathscr{S}_\OO \frac{\Omega_{\text{disc}}}{2\pi} + \vec{\mathscr{P}}_{\OO} \cdot \vec{b} + \nu n_{W}. 
\end{align}

So far, the topological action has been defined for flat gauge fields. This translates to the requirement that for any
region $W$, $\Phi_W/2\pi$, $\Omega_{\text{disc}}/2\pi$, $(1 - U (2\pi/M ))^{-1} \vec{b}$, and $n_W$ are integer-valued \cite{manjunath2021cgt}. Nevertheless, we can use the effective action to deduce the response to non-flat gauge field configurations, which corresponds to the presence of lattice disclinations and dislocations. This leads to the prediction:
\begin{align}
Q_W = \sigma_H \frac{\delta\Phi_{W,\OO}}{2\pi} + \mathscr{S}_\OO \frac{\Omega_{\text{disc}}}{2\pi} + \vec{\mathscr{P}}_{\OO} \cdot \vec{b}_\OO + \nu n_{W,\OO}  \mod q_\star .
\end{align}
Here we have incorporated the fact that in the presence of lattice disclinations, the Burgers vector $\vec{b}_\OO$ depends on $\OO$. We have also incorporated the fact that there may be fractional, irregular unit cells, and their counting empirically is $\OO$-dependent, leading to the $\OO$-dependence of $n_{W,\OO}$ and $\delta \Phi_{W,\OO}$. Finally, we take $Q_W$ modulo $q_\star$ because of the equivalences on $\mathscr{S}_{\OO}$ and $\vec{\mathscr{P}}_\OO$. Physically the $\mod q_\star$ arises because any local potential at the core of the defect can trap an anyon, changing the charge by a multiple of $q_\star$, and this is also beyond the topological field theory description. 

As explained in \cite{zhang2024pol}, we can treat corner angles and vectorized edge lengths on the same footing as disclination angles and Burgers vectors, so we can replace $\Omega_{\text{disc}}$ and $\vec{b}_\OO$ in the above equation with $\Gamma$ and $\vec{L}_\OO$ to get Eq.~\eqref{QwEq}.

\section{Parton effective field theory}\label{app:partonMF}

Here we use the parton effective field theory to derive the invariants of the FCI model wave functions in terms of the invariants of the parton states. 

Each parton state is defined by its own effective field theory. For the 1/2-Laughlin states, the full effective Lagrangian consists of the effective field theories for each parton state coupled to a dynamical $U(1)$ gauge field $\alpha$:

\begin{align}\label{eq:partonL}
    \nonumber\mathcal{L}&=-\frac{1}{4\pi}a\cup d a+\frac{1}{2\pi}(\alpha+ A+\mathscr{S}_{\OO,1}\omega+\vec{\mathscr{P}}_{\OO,1}\cdot\vec{R})\cup d a\\
    \nonumber&-\frac{1}{4\pi}b\cup d b+\frac{1}{2\pi}(-\alpha+\mathscr{S}_{\OO,2}\omega+\vec{\mathscr{P}}_{\OO,2}\cdot\vec{R})\cup d b.\\
\end{align}
Here $a, b$ are dynamical $\U$ gauge fields describing the two parton species whose field strength describes the conserved current of each parton. Integrating out $\alpha$ give us the constraint that $a=b$ up to a gauge transformation.
The effective field theory for $1/k$-Laughlin states can be similarly derived with $k$ parton flavors and $k-1$ internal $\U$ gauge field between them. Integrating out the internal gauge fields outputs the coupling:

\begin{align}\label{eq:parton_data}
    v &=1,
    \nonumber \\
    m &=\sum_{p}\kappa_{p}, 
    \nonumber \\
    \vec{t}_{\OO} &= [1-U(2\pi/M_{\OO})^T]\sum_{p}\vec{\mathscr{P}}_{\OO,p}
    \nonumber \\
    s_{\OO} &= \sum_{p}\mathscr{S}_{\OO,p},
    \nonumber \\
    \bar{\sigma}_H &=\mathscr{S}_{SPT,\OO}=\vec{\mathscr{P}}_{SPT,\OO}=\kappa_{SPT}=0.
\end{align}
Plugging into Eq.~\eqref{eq:invariants}, we have 

\begin{align}\label{eq:parton_data2}
    \mathscr{S}_{\OO} &= s_{\OO}/k = \frac{1}{k} \sum_{p=1}^k \mathscr{S}_{\OO,p},
    \nonumber \\
    \vec{\mathscr{P}}_{\OO} &= [1-U(2\pi/M_{\OO})^T]^{-1}\vec{t}_{\OO}/k = \frac{1}{k} \sum_{p=1}^k \vec{\mathscr{P}}_{\OO, p},
    \nonumber \\
    \kappa &= m/k = \frac{1}{k} \sum_{p=1}^k \kappa_p.
\end{align}

For invertible states, $\SO$ and $\PO$ transform under a shift of origin $\OO\rightarrow \OO+\vec{v}$ following Table ~\ref{table:shiftingO}. Note that every transformation rule is linear in $\SO$ and $\PO$. For the projected wave functions, Eqs.~\eqref{eq:parton_data}\eqref{eq:parton_data2} shows that $\SO$, $\vec{\mathscr{P}}_\OO$ is linear in the shift $\mathscr{S}_{\OO,p}$ and polarization $\vec{\mathscr{P}}_\OO$ of each parton state.  Therefore the transformations of Table ~\ref{table:shiftingO} are expected to hold for FCIs by linearity. Our numerical calculations confirm this.

\begin{figure*}[t]
    \centering
    \includegraphics[width=17.5cm]{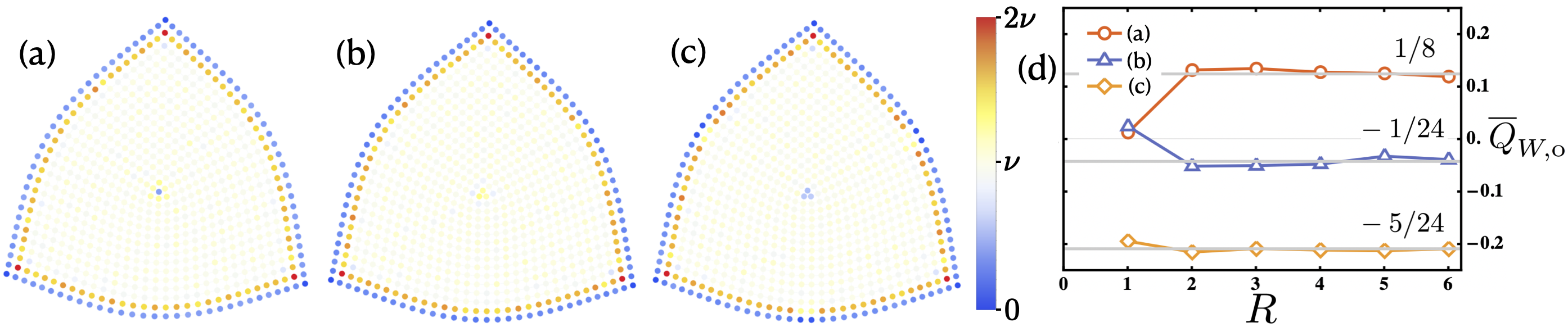}
    \caption{1/3-Laughlin state on a $\Omega=\pi/2$ disclination, filling 100 fermions in the whole disk.  \textbf{(a)} The charge profile in units of $\nu$ for $\OO=\beta$, $\mathscr{S}_{\beta,1}=\mathscr{S}_{\beta,2}=\mathscr{S}_{\beta,3}=1/2$; \textbf{(b)} $\OO=\alpha$, $\mathscr{S}_{\alpha,1}=1/2, \mathscr{S}_{\alpha,2}=\mathscr{S}_{\alpha,3}=7/2$; \textbf{(c)} $\OO=\alpha$, $\mathscr{S}_{\alpha,1}=1/2, \mathscr{S}_{\alpha,2}=\mathscr{S}_{\alpha,3}=5/2$. \textbf{(d)} $\overline{Q}_{W,\OO}$ as a function of radius of $W$.}
    \label{fig:disclination13}
\end{figure*}

\begin{figure*}[t]
    \centering
    \includegraphics[width=17.5cm]{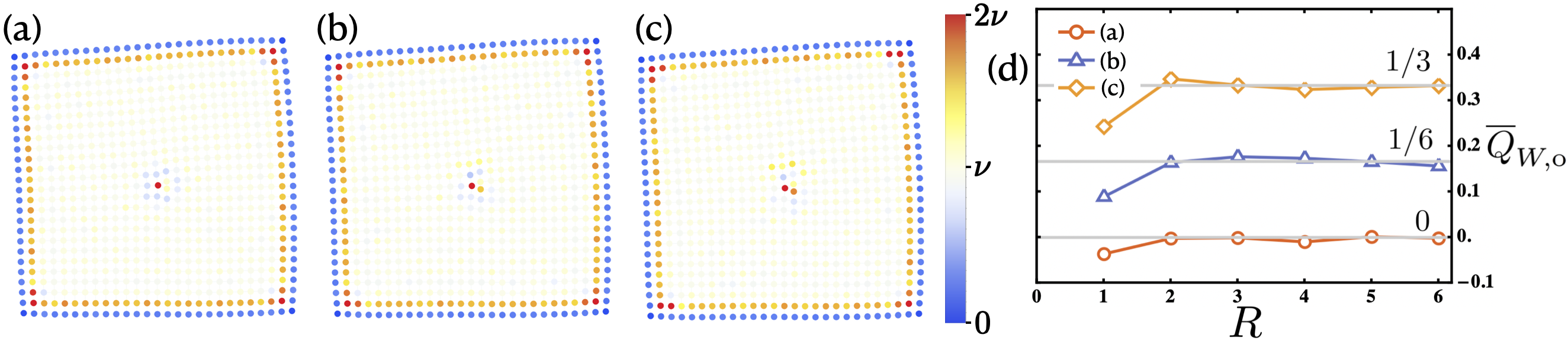}
    \caption{1/3-Laughlin state on a $\vec{b} = (0,1)$ dislocation, filling 90 fermions in the whole disk. \textbf{(a)} The charge profile in units of $\nu$ for $\OO=\alpha$, $\vec{\mathscr{P}}_{\alpha,1}=\vec{\mathscr{P}}_{\alpha,1}=\vec{\mathscr{P}}_{\alpha,1}=(0,0)$; \textbf{(b)} $\OO=\beta$, $\vec{\mathscr{P}}_{\beta,1}= \vec{\mathscr{P}}_{\beta,2}=(0,0), \vec{\mathscr{P}}_{\beta,3}=(1/2,1/2)$; \textbf{(c)} $\OO=\beta$, $\vec{\mathscr{P}}_{\beta,1} = (0,0)$, $ \vec{\mathscr{P}}_{\beta,2} = \vec{\mathscr{P}}_{\beta,3}=(1/2,1/2)$. \textbf{(d)} $\overline{Q}_{W,\OO}$ as a function of radius of $W$.}
    \label{fig:dislocation13}
\end{figure*}

\begin{figure*}[t]
    \centering
    \includegraphics[width=16cm]{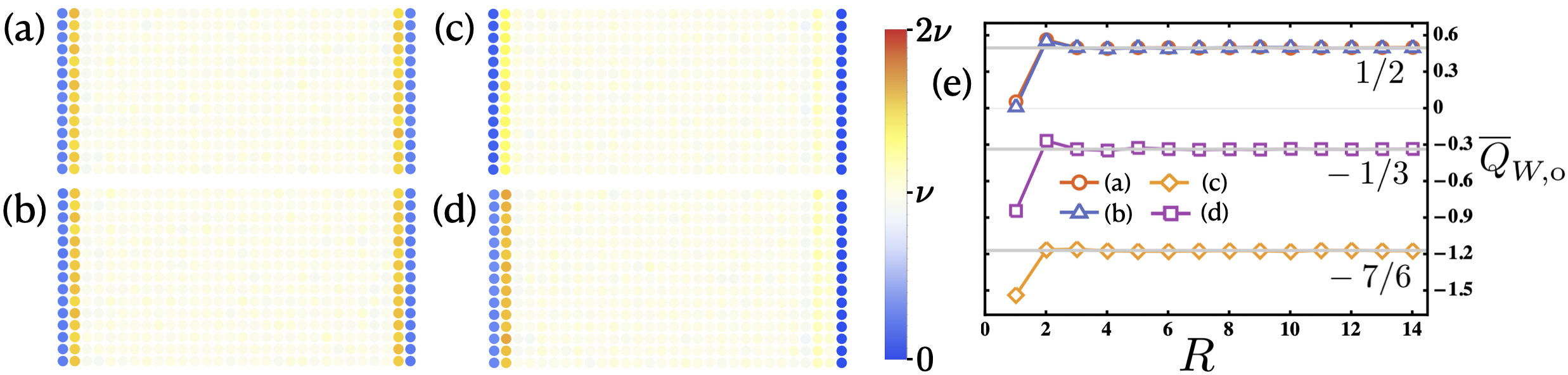}
    \caption{1/3-Laughlin state on cylinders, filling 60 fermions in the whole disk. We calculate the total charge on the left boundary. \textbf{(a)} The charge profile in units of $\nu$ with periodic boundary condition in the $y-$ direction for $\OO=\alpha$, $\vec{L_\OO}=(0,14)$, $\vec{\mathscr{P}}_{\alpha,1}=\vec{\mathscr{P}}_{\alpha,2}=\vec{\mathscr{P}}_{\alpha,3}=(0,0)$; \textbf{(b)} same as (a) but with $\vec{L_\OO}=(0,15)$; \textbf{(c)} $\OO=\beta$, $\vec{L_\OO}=(0,14)$,  $\vec{\mathscr{P}}_{\beta,1} = (1/2,1/2)$, $\vec{\mathscr{P}}_{\beta,2} = \vec{\mathscr{P}}_{\beta,3}=(0,0)$; \textbf{(d)} same as (c) but with $\vec{L_\OO}=(0,15)$
    \textbf{(e)} $\overline{Q}_{W,\OO}$ as a function of radius of $W$.}
    \label{fig:edge13}
\end{figure*}

\begin{figure*}[t]
    \centering
    \includegraphics[width=14.5cm]{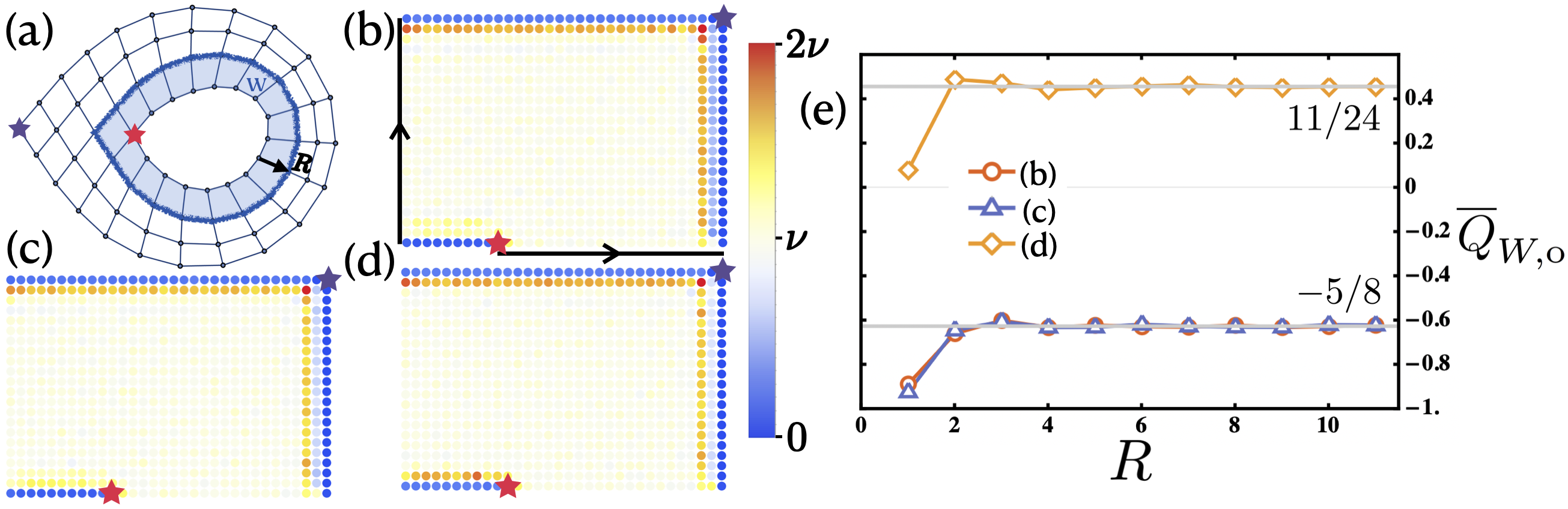}
    \caption{1/3-Laughlin state on a ribbon geometry, filling 80 fermions in the whole disk.  \textbf{(a)} The ribbon geometry, $W$ covers the inner coner. \textbf{(b)} The charge profile in units of $\nu$. We glue the two boundaries along the arrow direction to obtain the ribbon geometry, where the inner and outer corners are labeled with stars. Here $\OO=\beta$, $\Gamma=\frac{2\pi}{4}$, $\vec{L}_{\OO}\simeq(0,0)$, $\vec{\mathscr{P}}_{\beta,1}=\vec{\mathscr{P}}_{\beta,2}=\vec{\mathscr{P}}_{\beta,3}=(0,0)$, $\mathscr{S}_{\beta,1}=\mathscr{S}_{\beta,2}=\mathscr{S}_{\beta,3}=1/2$; \textbf{(c)} same as (b) but with $\vec{L}_{\OO}\simeq(0,1)$. \textbf{(d)} $\OO=\alpha$, $\vec{L}_{\OO}\simeq(0,0)$, $\vec{\mathscr{P}}_{\alpha,1}=\vec{\mathscr{P}}_{\alpha,2}=\vec{\mathscr{P}}_{\alpha,3}=(0,0)$, $\mathscr{S}_{\alpha,1}=\mathscr{S}_{\alpha,2}=1/2, \mathscr{S}_{\alpha,3}=7/2$. \textbf{(e)} $\overline{Q}_{W,\OO}$ as a function of radius of $W$. $W$ covers the inner boundary. $W$ covers the inner boundary.}
    \label{fig:ribbon13}
\end{figure*}

\section{Numerical Methods}\label{app:num_method}
\subsection{Constructing FCI wave function from ground states of Hofstadter model}

In this section, we review the construction of projected wave functions from the ground states of the free fermion Hofstadter model.  The Hofstadter model is defined by the Hamiltonian
\begin{equation}
    H = -t\sum_{<ij>} e^{-i A_{ij}} f_i^{\dagger} f_j + h.c.
\end{equation}
where $i,j$ are site indices, and $A_{ij}$ represents the vector potential that threads $\phi$ flux per unit cell.
The ground state is characterized by the topological invariants $\{c_-, C,\kappa, \Theta_{\OO}^\pm\}$ \cite{zhang2023complete}. It turns out that for the Hofstadter model, the many-body ground states with $C=\pm 1$ can be uniquely labeled by $C,\kappa$ (though this is generally not true for states with $|C|>1$); we label these $C = \pm 1$ states as $\ket{\psi_{C,\kappa}}$. For these $C = \pm 1$ states, once $C$ and $\kappa$ are specified, the corresponding Hofstadter lobes can be identified, and the quantities $\SO$ and $\PO$ can be determined as described in \cite{zhang2022fractional,zhang2022pol}. We can then define different flavors of parton ground states labeled by $p$, $\{\ket{\psi_{C_p,\kappa_p}}\}$.
These parton ground states must be defined on the same lattice and have the same filling, but they need not share the same topological invariants. 

The FCI wave function is then created by projecting these parton ground states. The 1/2-Laughlin states require two parton flavors with $C_1 = C_2 = 1$, and their wave functions are constructed by projecting the partons to the same locations:
\begin{align}
    \Psi_{1/2}(\{r_i\}) = \psi_{C_1, \kappa_1}(\{r_i\}) \psi_{C_2, \kappa_2}(\{r_i\}) 
\end{align}
The 1/3-Laughlin states require three parton flavors with $C_1 = C_2 = C_3 = 1$, and their wave functions are written as
\begin{align}
    \Psi_{1/3}(\{r_i\}) = \psi_{C_1, \kappa_1}(\{r_i\}) \psi_{C_2, \kappa_2}(\{r_i\}) \psi_{C_3, \kappa_3}(\{r_i\}) 
\end{align}

The ground states $\psi_{C, \kappa}(\{r_i\})$ with $N$ particles are Slater determinant states:
\begin{align}
    \psi_{C, \kappa}(\{r_i\}) = \frac{1}{\sqrt{N!}}\text{Det}(\mathcal{M}),
\end{align}
where 
\begin{align}
\mathcal{M}_{ij}=\chi_{i}(r_j),
\end{align}
and $\chi_{i}$ are the single particle eigenstates.

\subsection{Variational Monte Carlo}

In this section, we outline the process of sampling the projected wave function $\ket{\Psi}$ from parton construction using the Metropolis Monte Carlo method.

We aim to calculate the charge at each site $Q_i=\bra{\Psi}\hat{n}_{i}\ket{\Psi}$, where $\hat{n}_{i}$ is the number operator. The expectation value $Q_i$ can be expressed as

\begin{align}\label{eq:expectationVal}
    \nonumber &\bra{\Psi}\hat{n}_{i}\ket{\Psi}\\
    \nonumber
    =& \sum_{r_1,r_2,\dots r_{N}} \langle \Psi | \hat{n}_{i} | r_1,r_2,\dots r_{N} \rangle \langle r_1,r_2,\dots r_{N} | \Psi \rangle\\
    \nonumber=&\sum_{r_1,r_2,\dots r_{N}} \frac{\bra{\Psi}\hat{n}_i\ket{r_1,r_2,\dots r_{N}}}{\bra{\Psi}\ket{r_1,r_2,\dots r_{N}}}P(r_1,r_2,\dots r_{N}),\\
    =&\sum_{r_1,r_2,\dots r_{N}} \delta_i P(r_1,r_2,\dots r_{N})
\end{align}
where $P(r_1,r_2,\dots r_{N}):= |\bra{\Psi}\ket{r_1,r_2,\dots r_{N}}|^2$ is the probability of the configuration $\ket{r_1,r_2,\dots r_{N}}$, and $\delta_i$ is defined as

\begin{align}
    \delta_i=\begin{cases}
        1  &\text{if  } r_i\in \{r_1,r_2,\dots r_N\}\\
        0  &\text{otherwise}
    \end{cases}
\end{align}

This expectation value $Q_i$ is sampled using the Metropolis-Hastings algorithm \cite{hastings2010measuring} through the following steps:

\begin{enumerate}
    \item Start with a random configuration $\ket{r_1,r_2,\dots r_{N}}$
    \item \textit{Propose} a new configuration $\ket{\tilde{r}_1,\tilde{r}_2,\dots \tilde{r}_{N}}$ by moving one particle to a random unfilled position. \textit{Update} to this new configuration if 
    \begin{align}
        \nonumber \left|\frac{\bra{\Psi}\ket{\tilde{r}_1,\tilde{r}_2\dots \tilde{r}_{N}}}{\bra{\Psi}\ket{r_1,r_2,\dots r_{N}}}\right|^2>q,
    \end{align}
    where $q$ is a random real number between 0 to 1 ($q$ is chosen randomly at each step). Otherwise \textit{Reject} the proposal and revert to the old configuration $\ket{r_1,r_2,\dots r_{N}}$ if the above check is false.
    \item Repeat Step 2. The expectation value of $Q_i$ in Eq.~\eqref{eq:expectationVal} is obtained as the average of $\delta_i$ over the sample trajectory (with the first few iterations discarded as they do not converge to the target distribution).
\end{enumerate}

\section{Extra numerical data}\label{app:num}

In the main text, we have shown numerical data
for the $1/2$ Laughlin state on the square lattice and $1/3$ Laughlin state on the honeycomb lattice. In both of these cases, it is possible to have non-trivial symmetry fractionalization as specified by the spin vector $\vec{s}_\OO$ and discrete torsion vector $\vec{t}_{\OO}$ reviewed above. 

In this section, we consider the 1/3-Laughlin states on the square lattice, for which $\vec{s}_{\OO}$ and $\vec{t}_{\OO}$ is necessarily trivial according to the symmetry fractionalization classification \cite{barkeshli2019,manjunath2020FQH,bulmashSymmFrac}. This implies that $\overline{Q}_{W,\OO} \mod q_\star$ only receives contributions from the SPT terms in the topological field theory:
\begin{align}\label{eq:trivialC}
    \nonumber\overline{Q}_{W,\OO}&=  \vec{L}_{\OO}\cdot \vec{\mathscr{P}}_{SPT,\OO}\\
    &+\frac{ \Gamma\mathscr{S}_{SPT,\OO}}{2\pi} \mod q_\star
\end{align}

We verify this by calculating charge response for disclinations, dislocations, edges and ribbons as in the main text. The results are shown in Fig. \ref{fig:disclination13}, \ref{fig:dislocation13}, \ref{fig:edge13}, \ref{fig:ribbon13}, all of which agree with Eq.~\eqref{eq:trivialC}.

\bibliography{bibliography}
\clearpage
\end{document}